\documentstyle[prd,aps,floats,tighten,epsfig,graphicx,color]{revtex}

\newcommand{\beq}{\begin{equation}}
\newcommand{\eeq}{\end{equation}}
\newcommand{\bea}{\begin{eqnarray}}
\newcommand{\ena}{\end{eqnarray}}
\newcommand{\etal}{{\it et al.}}
\newcommand{\ie}{{\it i.e.}}
\newcommand{\lsim}{\mathrel{\mathop{\kern 0pt \rlap
{\raise.2ex\hbox{$<$}}}
\lower.9ex\hbox{\kern-.190em $\sim$}}}
\newcommand{\gsim}{\mathrel{\mathop{\kern 0pt \rlap
{\raise.2ex\hbox{$>$}}}
\lower.9ex\hbox{\kern-.190em $\sim$}}}

\newcommand{\app}[3]{Astropart.\ Phys.\ {\bf #1}, #3 (#2)}

\newcommand{\hepph}[1]{{\tt hep-ph/#1}}
\newcommand{\astroph}[1]{{\tt astro-ph/#1}}

\newcommand{\plb}[3]{Phys.\ Lett.\ B\ {\bf #1}, #3 (#2)}
\newcommand{\npb}[3]{Nucl.\ Phys.\ B\ {\bf #1}, #3 (#2)}

\renewcommand{\apj}[3]{Astrophys.\ J.\ {\bf #1}, #3 (#2)}

\renewcommand{\prl}[3]{Phys.\ Rev.\ Lett. {\bf #1}, #3 (#2)}
\renewcommand{\prd}[3]{Phys.\ Rev.\ D\ {\bf #1}, #3 (#2)}

\newcommand{\href}[2]{#1}

\newcommand{\rclose}{\mbox{$\rho_{C}^{\, 0}$}}
\newcommand{\omegaM}{\mbox{$\Omega_{\rm m}$}}
\newcommand{\omegaB}{\mbox{$\Omega_{\rm b}$}}
\newcommand{\omegaL}{\mbox{$\Omega_{\Lambda}$}}

\newcommand{\omegachi}{\mbox{$\Omega_{\chi}$}}
\newcommand{\Ochi}{\mbox{$\Omega_{\chi} \, h^{2}$}}
\newcommand{\etaphi}{\mbox{$\eta_{\Phi}$}}
\newcommand{\yF}{\mbox{$y_{F}$}}
\newcommand{\xF}{\mbox{$x_{F}$}}
\newcommand{\TF}{\mbox{$T_{F}$}}
\newcommand{\Sa}{\mbox{$\sigma_{\rm an}$}}
\newcommand{\nA}{\mbox{$n_{\chi}$}}
\newcommand{\nAe}{\mbox{$n_{\chi}^{\, 0}$}}
\newcommand{\fA}{\mbox{$f_{\chi}$}}
\newcommand{\fF}{\mbox{$f_{F}$}}
\newcommand{\FA}{\mbox{$F_{\chi}$}}
\newcommand{\fAe}{\mbox{$f_{\chi}^{\, 0}$}}
\newcommand{\fAasy}{\mbox{$f_{\chi}^{\, \rm asy}$}}

\definecolor{cyan}{cmyk}{1.,0.,0.,0.5}
\definecolor{magenta}{cmyk}{0.,1.,0.,0.5}
\definecolor{verdatre}{cmyk}{0.5,0.,0.5,0.5}
\definecolor{yellow}{cmyk}{0.,0.,0.2,0.0}
\definecolor{rouge}{cmyk}{0.,0.4,0.6,0.0}
\definecolor{orange}{cmyk}{0.,0.5,0.5,0.}
\definecolor{violet}{rgb}{0.5,0.,0.5}
\definecolor{red}{rgb}{0.0,0.0,0.0}

\begin{document}

\draft
\title{Quintessence and the Relic Density of Neutralinos}
\vskip 1.cm
\author{
Pierre Salati$^{\rm a,b}$
\footnote{E--mail: salati@lapp.in2p3.fr}}
\vskip 0.5cm
\address{
\begin{flushleft}
a) Laboratoire de Physique Th\'eorique LAPTH, B.P.~110, F-74941
Annecy-le-Vieux Cedex, France.\\
b) Universit\'e de Savoie, B.P.~1104, F-73011 Chamb\'ery Cedex,
France.
\end{flushleft}
}
\maketitle
\vskip 0.5cm
\textcolor{red}{
\centerline{July 17, 2003 -- LAPTH--951/02}}

\vskip 0.5cm
\begin{abstract}
The archetypal model for the recently discovered dark energy component
of the universe is based on the existence of a scalar field whose
dynamical evolution comes down today to a non--vanishing cosmological
constant. In the past -- before big--bang nucleosynthesis
\textcolor{red}{for that matter}
-- that scalar field could have gone through a period of kination during
which the universe has expanded at a much higher pace than what is
currently postulated in the standard radiation dominated cosmology.
I~examine here the consequences of such a period of kination on the
relic abundance of neutralinos and find that the latter could be much higher
-- by three orders of magnitude -- than what is estimated in the canonical
derivation. I~shortly discuss the implications of this scenario for
the dark matter candidates and their astrophysical signatures.
\end{abstract}
\vskip 1.cm

\section{Introduction.}
\label{sec:introduction}

\vskip 0.1cm
The
\textcolor{red}{recent WMAP}
observations of the Cosmic Microwave Background (CMB) anisotropies
\cite{WMAP}, combined either with the determination of the relation
between the distance of luminosity and the redshift of supernovae SNeIa
\cite{supernovae_omegaL}, or with the large scale structure (LSS)
information from galaxy and cluster surveys \cite{2dF}, give
independent evidence for a cosmological average
\textcolor{red}{
matter density of $\omegaM = 0.27 \pm 0.04$} \cite{WMAP}.
This value may be compared to a baryon density of
\textcolor{red}{
$\omegaB = 0.044 \pm 0.004$} as indicated by nucleosynthesis
\cite{nucleosynthesis} and the relative heights of the first acoustic
peaks in the CMB data. A significant fraction of the matter in the
universe is dark and non--baryonic.
The cosmological observations also point towards a flat universe
\textcolor{red}{with $\Omega_{\rm tot} = 1.02 \pm 0.02$} and
strongly favour the existence of a cosmological constant which contributes
a fraction
\textcolor{red}{$\omegaL = 0.73 \pm 0.04$} to the closure density.
The pressure--to--density ratio $w$ of that fluid is negative with a value
of $w = - 1$ in the case of an exact cosmological constant. That $\omegaL$
component is called dark energy as opposed to the $\omegaM$ dark matter
contribution.

\vskip 0.1cm
The nature of the astronomical dark matter is still unresolved insofar.
The favorite candidate for the non--baryonic component is a weakly--interacting
massive particle (WIMP). The so--called neutralino naturally arises in the
framework of supersymmetric theories.
Large efforts have been devoted in the past decade to pin down these evading
species. New experimental techniques have been devised to look for the direct
and indirect astrophysical signatures of the presence of neutralinos in our
Milky--Way \cite{gamma_neutralino_MW} as well as in extra--galactic systems
\cite{gamma_neutralino_M87}. The uncertainty on the theoretical estimates of
the various signals has been considerably reduced. As an illustration, the
energy spectrum of secondary spallation antiprotons -- the natural background
to a putative neutralino--induced antiproton extra radiation -- is now well
under control \cite{secondary_antiprotons}.
Another example of the level of sophistication which the theoretical
investigations have reached is provided by the calculations of the neutralino
relic abundance $\omegachi$. The observation that this relic density -- depending
on the numerous parameters of the model -- falls in the ballpark of the measured
value for $\omegaM$ has been a crucial argument in favor of supersymmetric
particles as a viable option to non--baryonic dark matter. A large number of
processes -- typically $\sim$ 2000 -- are now taken into account and the
corresponding diagrams are automatically generated and calculated with the help
of numerical codes such as micrOMEGAs \cite{micromegas}. Co--annihilations are
taken into account and the thermal averaging $\left< \sigma v \right>$ of the
product of the velocity by the cross section is performed.

\vskip 0.1cm
Surprisingly enough, calculations of $\omegachi$ are based on the assumption
that the universe is dominated by radiation when neutralinos decouple from
the primordial plasma and reach their relic density. This hypothesis is
presumably correct as soon as primordial nucleosynthesis (BBN) sets in
at a time of $\sim$ 1 second. We have however little
information on the earlier pre--BBN period, a crucial stage during which
neutralinos freeze out. If the expansion of the universe is modified with
respect to a pure radiation--dominated behaviour, the quenching of these
species could be drastically modified. An increase in the expansion rate $H$
accelerates the  decoupling of neutralinos and translates into larger values
for the relic density $\omegachi$.

\vskip 0.1cm
Exploring the effects of a modified expansion history of the universe
onto the relic abundance of neutralinos is no longer a mere academic
exercise. Such an analysis has become mandatory inasmuch as a new and
unexpected component -- the dark energy $\omegaL$ -- has been discovered.
The potential interplay between that component and its matter counterpart
$\omegaM$ is worth being explored and may have unexpected consequences.
The archetypal model for the cosmological dark energy is the so--called
quintessence
\textcolor{red}{\cite{quintessence,steinhardt}} and relies on the existence
of a neutral scalar field $\Phi$ with Lagrangian density
\beq
{\cal L} \; = \; \frac{1}{2} \, g^{\, \mu \nu} \,
\partial_{\mu}  \Phi \, \partial_{\nu}  \Phi
\, - \, V \left( \Phi \right) \;\; .
\label{scalar_neutral}
\eeq
Should the field $\Phi$ be homogeneous and the metric be flat, the energy
density may be expressed as
\beq
\rho_{\Phi} \equiv T^{0}_{\; 0} \; = \;
{\displaystyle \frac{\dot{\Phi}^{2}}{2}} \, + \, V \left( \Phi \right) \;\; ,
\label{energy_density}
\eeq
whereas the pressure obtains from $T_{i j} \equiv - g_{\, i j} \, P$
so that
\beq
P_{\Phi} \; = \; {\displaystyle \frac{\dot{\Phi}^{2}}{2}} \, - \, V \left( \Phi \right)
\;\; .
\label{pressure}
\eeq
If the kinetic term ${\dot{\Phi}^{2}}/{2}$ is negligible with respect to the
contribution of the potential $V \left( \Phi \right)$, a pure cosmological
constant with $w_{\Phi} = P_{\Phi} / \rho_{\Phi} = - 1$ is recovered since
$\rho_{\Phi} = - \, P_{\Phi} = V \left( \Phi \right)$. As indicated by
cosmological observations, this is the case today.
\textcolor{red}{
But the field $\Phi$ has been continuously rolling down. Should the
kinetic term ${\dot{\Phi}^{2}}/{2}$ have dominated over the potential
$V \left( \Phi \right)$ in the early universe, a period of kination
-- \ie, domination by the kinetic energy of the field $\Phi$ --
would have ensued with drastic effects on the expansion rate of the
universe \cite{joyce_kination}.}

\vskip 0.1cm
\textcolor{red}{
In section~\ref{sec:kination}, we briefly recall why a pure cosmological
constant should be disregarded and replaced by a dynamical dark
energy component in the form of a scalar field, the so--called
quintessence whose salient features are presented. The existence
of tracking solutions provides a natural solution to the problem of
initial conditions. We also pay some attention to the difficulty
of generating a kination--dominated expansion in the early universe
together with a cosmological constant today \cite{steinhardt}. We
show that this difficulty may be circumvented depending on the
potential $V \left( \Phi \right)$ that drives the evolution of the
scalar field and we propose examples where quintessence boosts the
expansion rate in the past while it still accounts for $\omegaL$ today.}
Following a suggestion by \cite{joyce_idea}, we investigate in
section~\ref{sec:decoupling} the effects of kination on the thermal
decoupling of neutralinos and derive an approximate relation between
their relic abundance $\omegachi$ and their annihilation cross section
in the presence of kination. Section~\ref{sec:discussion} is devoted to
a discussion of the consequences of this scenario on the astrophysical
signatures of neutralino dark matter.

\section{Kination and the expansion rate of the universe.}
\label{sec:kination}

\vskip 0.1cm
\textcolor{red}{
Two difficulties arise with a pure cosmological constant.
The coincidence or fine--tuning problem lies in the fact
that the vacuum energy $\rho_{\Lambda}^{0}$ comes into
play only today and is therefore of order the closure density
$\rclose$, an exceedingly small value with respect to the typical
Planck energy scale $M_{\rm Planck}^{4}$ set by particle
physics
\beq
{\displaystyle \frac{\rho_{\Lambda}^{0}}{M_{\rm Planck}^{4}}} \sim
{\displaystyle \frac{\rm 4 \; keV \; cm^{-3}}
{\left\{ \rm 1.22 \times 10^{19} \; GeV \right\}^{4}}}
\sim 10^{- 123} \;\; .
\eeq
The other issue is related to the initial conditions in which
a pure cosmological constant has to be prepared. At the Planck
time, the corresponding vacuum energy density
$\rho_{\Lambda}^{i} = \rho_{\Lambda}^{0}$ needs to be exceedingly
fine--tuned with respect to the radiation density
\beq
{\displaystyle \frac{\rho_{\Lambda}^{i}}{\rho_{\rm rad}^{i}}} \sim
{\displaystyle \frac{\rho_{\Lambda}^{0}}{\rho_{\rm rad}^{0}}} \;
\left\{ {\displaystyle \frac{T_{0}}{M_{\rm Planck}}} \right\}^{4}
\sim 10^{- 125} \;\; .
\eeq
Because a pure cosmological constant does not vary in time, similar
values are obtained in the previous relations. This has led to some
confusion between the fine--tuning and the initial condition problems.
Quintessence actually solves only the latter difficulty while the
fine--tuning of $\rho_{\Lambda}^{0}$ with respect to $M_{\rm Planck}^{4}$
is forced by direct observation.}

\vskip 0.1cm
\textcolor{red}{
The idea of quintessence is based on the existence of a scalar field
$\Phi$ that rolls down its potential $V$ according to the homogeneous
Klein--Gordon equation
\beq
\ddot{\Phi} \, + \,3 \, H \, \dot{\Phi} \, + \,
{\displaystyle \frac{\partial V}{\partial \Phi}} \; = \; 0 \;\; .
\label{KG_1}
\eeq
Should $\Phi$ satisfy to that relation, its energy--momentum tensor
would be immediately conserved as may be inferred from the identity
\beq
D_{\mu} T^{\, \mu \alpha} \; = \;
\left( \partial^{\, \alpha} \varphi \right) \, \cdot \, \left\{
D_{\mu} \left( \partial^{\, \mu} \varphi \right) \, + \,
{\displaystyle \frac{dV}{d \varphi}} \right\} \;\; .
\label{relation_TIE_KG}
\eeq
That energy--momentum tensor obtains from the Lagrangian
density (\ref{scalar_neutral}) and may be written as
\beq
T_{\, \mu \nu} \; = \;
\partial_{\mu} \varphi \, \partial_{\nu} \varphi \, - \,
g_{\mu \nu} \, {\cal L} \;\; .
\label{impulsion_energie}
\eeq
Its conservation translates into the adiabatic expansion of
quintessence as
\beq
{\displaystyle \frac{d}{dt}} \,
\left( \rho_{\Phi} \, a^{3} \, \right) \; = \; - \, P_{\Phi} \,
{\displaystyle \frac{d a^{3}}{dt}} \;\; ,
\label{KG_2}
\eeq
where the dark energy density $\rho_{\Phi}$ and pressure $P_{\Phi}$
have been defined in relations~(\ref{energy_density}) and (\ref{pressure}).
The expansion rate $H = \dot{a}/a$ is increased by the presence
of quintessence
\beq
H^{2} \; = \; {\displaystyle \frac{8 \, \pi \, G}{3}} \,
\left\{ \rho_{\rm B} \, + \, \rho_{\Phi} \right\} \;\; ,
\eeq
where the energy density of the background
$\rho_{\rm B} = \rho_{\rm rad} + \rho_{\rm m}$ is
dominated by radiation in the past and by matter since equality.
Notice that at fixed equation--of--state $w = P / \rho$, the
energy density scales as
\beq
\rho \propto
a^{\displaystyle - \, 3 \, \left( 1 + w \right)} \;\; .
\eeq
As discussed in \cite{steinhardt}, we are looking for tracking solutions
of (\ref{KG_1}) for which $w_{\rm B} \ge w_{\Phi} \simeq {\rm constant}$.
This allows a natural and smooth decrease of the quintessence energy density
$\rho_{\Phi}$ that remains subdominant until recently when it takes over
from the background $\rho_{\rm B}$. This scenario is realized only if
$\left| {V'}/{V} \right|$ decreases with $V$ as time goes on. This
translates into the condition
\beq
\Gamma = {\displaystyle \frac{V'' V}{V'^{\, 2}}} > 1 \;\; ,
\eeq
where the prime denotes the derivative with respect to the field $\Phi$.
As long as the background dominates
-- for $\Omega_{\Phi} = \rho_{\Phi} / ( \rho_{\Phi} + \rho_{\rm B} ) \ll 1$ --
the parameter $\Gamma$ is related to the equation--of--state $w_{\Phi}$
of the tracking solution through
\beq
w_{\Phi} \; = \;
{\displaystyle \frac{w_{\rm B} \, - \, 2 \, \left( \Gamma - 1 \right)}
{1 \, + \, 2 \, \left( \Gamma - 1 \right)}} \;\; .
\eeq
Inverse power law potentials $V \propto \Phi^{- \alpha}$ with $\alpha > 0$
naturally drive quintessence since $\Gamma - 1 = 1 / \alpha > 0$. On the
contrary, exponential potentials
$V \propto \exp \left( - \Phi / M \right)$ for which $\Gamma = 1$
and $w_{\Phi} = w_{\rm B}$ should be disregarded.}

\vskip 0.1cm
\textcolor{red}{
As shown in \cite{steinhardt}, the tracking solutions are attractors
towards which the field $\Phi$ relaxes for a large range of values of
the initial energy density $\rho_{\Phi}^{\rm i}$. Quintessence solves
therefore the initial condition problem to a large extent -- actually
to the extent that $\rho_{\Phi}^{\rm i}$ is comprised between
$\rho_{\Lambda}^{0}$ and $\rho_{\rm B}^{\rm i}$. The Klein--Gordon
equation (\ref{KG_1}) may be written as
\beq
\left| {\displaystyle \frac{V'}{V}} \right| \; = \; 3 \,
\sqrt{\displaystyle \frac{\kappa}{\Omega_{\Phi}}} \,
\sqrt{1 + w_{\Phi}} \, \left\{
1 \, + \, \frac{1}{6} {\displaystyle \frac{d \ln x}{d \ln a}} \right\} \;\; ,
\label{KG_3}
\eeq
where $\kappa = 8 \, \pi \, G \, / \, 3$ and
\beq
x \; = \; {\displaystyle \frac{\dot{\Phi}^{2} / 2}{V}} \; = \;
{\displaystyle \frac{1 + w_{\Phi}}{1 - w_{\Phi}}} \;\; .
\eeq}

\newpage
\textcolor{red}{
\noindent
Along the tracking solution, $w_{\Phi}$ and $x$ are constant.
When quintessence pops above the background
-- when $\rho_{\Phi} \sim \rho_{\rm B}$ -- relation~(\ref{KG_3})
translates into
\beq
{\displaystyle \frac{\Phi}{M_{\rm Planck}}} \simeq
{\displaystyle \frac{\alpha}{\sqrt{24 \, \pi}}} \,
\sqrt{\displaystyle \frac{\Omega_{\Phi}}{1 + w_{\Phi}}}
\sim {\cal O}(1) \;\; ,
\eeq
in the case of an inverse power law potential.
The field $\Phi$ is naturally of order the Planck mass $M_{\rm Planck}$
when quintessence takes over from the background. The actual coincidence
-- forced by direct observation -- lies in the fact that quintessence
becomes dominant today. As its energy density
-- and therefore the potential $V$ -- are of order the closure density
$\rclose$, this implies a significant amount of fine--tuning as
\beq
V \left( \Phi_{0} \right) \; = \; \mu^{4} \, \left\{
{\displaystyle \frac{M_{\rm Planck}}{\Phi_{0}}} \right\}^{\alpha}
\simeq \mu^{4} \sim \rclose \simeq 10^{- 47} \, {\rm GeV^{4}} \;\; ,
\eeq
where $\Phi_{0} \sim M_{\rm Planck}$ is the present value of the scalar
field. We conclude at this stage that quintessence does not solve the
coincidence or fine--tuning problem \cite{copeland}. This point cannot
be too strongly emphasized.}

\vskip 0.1cm
\textcolor{red}{
As also sketched in \cite{steinhardt}, the entire scenario of quintessence
would be in jeopardy should the initial energy density $\rho_{\Phi}^{\rm i}$
exceed $\rho_{\rm B}^{\rm i}$. If so, the initial position from which the
scalar field starts to roll lies too high on the potential. The field
$\Phi$ falls precipitously. As its kinetic energy becomes dominant, a kination
stage ensues during which the potential $V$ plays no role. The Hubble friction
-- see the Klein--Gordon equation (\ref{KG_1}) -- slows down the fall and
the field gets frozen at the value $\Phi_{\rm F}$ such that
\beq
{\displaystyle \frac{\Phi_{\rm F}}{M_{\rm Planck}}} \simeq
\sqrt{\displaystyle \frac{3}{4 \, \pi}} \, \left\{
1 \, + \, \frac{1}{2} \ln \Omega_{\rm i} \right\} \;\; ,
\eeq
where
$\Omega_{\rm i} = \rho_{\Phi}^{\rm i} / \rho_{\rm B}^{\rm i}$.
If, in the early universe, quintessence drives a period of kination 
with $\Omega_{\rm i} \gg 1$, the field gets frozen at
$\Phi_{\rm F} \sim 10 - 100 \; M_{\rm Planck}$, a value that exceeds
by far what is required by the conventional quintessence scheme where
$\Phi_{0} \sim M_{\rm Planck}$. The field has overshot the tracking
solution and its energy density has become so small
-- the potential $V$ decreases with increasing $\Phi$ -- that it
cannot account for $\rho_{\Lambda}^{0}$. A period of early kination
seems therefore incompatible with a near constant dark energy
density that becomes dominant today.}

\vskip 0.1cm
\textcolor{red}{
The problem raised by overshooting lies in the incompatibility between
the conditions
$\Phi_{\rm F} \gg M_{\rm Planck}$ and $\Phi_{0} \sim M_{\rm Planck}$.
However the latter has been derived in the framework of inverse power law
potentials. We anticipate that the overshooting difficulty could be
circumvented with other potentials.
%
As an illustration, we consider the toy--model where
\beq
V \left( \Phi \right) \; = \; \mu^{4} \;
\exp \left\{ {\displaystyle \frac{M}{\Phi}} \right\} \;\; ,
\label{counter_example}
\eeq
with $M \sim {\cal O} \left( M_{\rm Planck} \right)$. In order for
dark energy to overcome the background only today, we still need
-- as before -- a considerable amount of fine--tuning as
$\mu^{4} \sim \rclose \simeq 10^{- 47} \, {\rm GeV^{4}}$.
That potential has nevertheless the correct behaviour with
$\Gamma = {\displaystyle {V'' V}/{V'^{\, 2}}} =
1 \, + \, {\displaystyle {2 \, \Phi}/{M}} > 1$ and an
associated equation--of--state
\beq
w_{\Phi} \, = \, - \,
\left\{ 1 \, + \, {\displaystyle \frac{M}{4 \, \Phi}} \right\}^{-1}
\;\; ,
\eeq
on the tracking solution. The initial condition problem is solved
to the extent that $\rho_{\Phi}^{\rm i}$ is now larger than
$\rho_{\rm B}^{\rm i}$ and consequently drives a period of kination.
In that case, the field rapidly freezes at
$\Phi_{\rm F} \gg M_{\rm Planck} \sim M$
-- irrespective of its precise initial value --
and the potential reaches an asymptotic value of
\beq
V \left( \Phi_{\rm F} \gg M \right) \simeq
\mu^{4} \; = \; \rho_{\Lambda}^{0} \;\; .
\eeq
This example is remarkable as early kination is now mandatory
in order to get today a cosmological constant with
$w_{\Phi} \simeq - \, 1$.}

\newpage
\textcolor{red}{
In our counter--example to the overshooting no go theorem,
the potential~(\ref{counter_example}) has a sharp decrease for small
values of the field and then varies smoothly for $\Phi \gg M_{\rm Planck}$.
A dynamical realization of that idea -- proposed in \cite{rosati} --
is based on
\beq
V \left\{ \Phi_{1} , \Phi_{2} \right\} \; = \;
M^{\alpha + 4} \,
\left\{ \Phi_{1} \Phi_{2} \right\}^{- \alpha / 2} \;\; ,
\eeq
where two scalar fields come into play. If one of the fields starts not
too far from the tracking solution whereas the other one is dropped from
an elevated position on the potential, we still get kination in the past
whereas a cosmological constant is recovered today. This behaviour occurs
for a wide range of initial conditions on $\Phi_{1}$ and $\Phi_{2}$.
Other examples are presented in \cite{rosati} with the same trend and
we conclude at this stage that a period of early dominant kination
could perfectly occur at the time of neutralino decoupling without
precluding a significant contribution today from a near constant dark
energy density.}

\vskip 0.1cm
\textcolor{red}{
In the standard big bang model, the early universe is only filled with
a gas of ultra--relativistic particles that behaves like a radiation
of photons. If the kinetic energy ${\dot{\Phi}^{2}}/{2}$ of some
additional scalar field component dominates both the potential
$V \left( \Phi \right)$ and the radiation energy density
$\rho_{\rm rad}$, a period of kination sets in. The overall energy
density decreases consequently like
\beq
\rho_{\rm tot} \simeq \rho_{\Phi} \simeq
{\displaystyle \frac{\dot{\Phi}^{2}}{2}} \, \propto \,
a^{-6} \;\; ,
\eeq
with respect to the scale factor $a$. This amounts to say that the derivative
${\partial V}/{\partial \Phi}$ is negligible with respect to the damping
term $3 \, H \, \dot{\Phi}$ in the Klein--Gordon equation (\ref{KG_1})
as the field tumbles down the potential.}
%
The energy density of radiation is related to the temperature $T$
of the primordial plasma through
\beq
\rho_{\rm rad} \; = \; g_{\rm eff}(T) \,
{\displaystyle \frac{\pi^{2}}{15}} \, T^{4} \;\; .
\eeq
The effective number $g_{\rm eff}$ of degrees of freedom allows
to express $\rho_{\rm rad}$ in units of the corresponding photon
density $\rho_{\gamma}$. The evolution of the scale factor $a$
with respect to the temperature $T$ follows from the requirement
that the entropy of the radiation is conserved so that
\beq
S_{\gamma} \, \propto \,
{\displaystyle \frac{4 \, \pi^{2}}{45}} \, h_{\rm eff}(T) \,
T^{3} \, a^{3}
\eeq
remains constant as $T$ drops. The effective number $h_{\rm eff}$
of entropic degrees of freedom is very close to $g_{\rm eff}$ since
\beq
h_{\rm eff} \, \simeq \, g_{\rm eff} \, \simeq \,
{\displaystyle \sum_{\rm B}} \,
{\displaystyle \frac{g_{\rm B}}{2}} \; + \;
{\displaystyle \sum_{\rm F}} \,
{\displaystyle \frac{7 \, g_{\rm F}}{16}} \;\; .
\eeq
The sum is carried over the bosonic $g_{\rm B}$ and fermionic
$g_{\rm F}$ spin states that correspond to those species that
are massless at temperature $T$. We infer that at any given time,
the scale factor $a$ and the temperature $T$ are related through
\beq
{\displaystyle \frac{a}{a_{0}}} \; = \;
\left\{
{\displaystyle \frac{h_{\rm eff}(T_{0})}{h_{\rm eff}(T)}}
\right\}^{1/3} \,
{\displaystyle \frac{T_{0}}{T}} \;\; ,
\label{a_to_T}
\eeq
where $a_{0}$ is the scale factor at some temperature of reference
$T_{0}$. During the decoupling of neutralinos, $g_{\rm eff}$ and
$h_{\rm eff}$ do not change drastically so that $a$ varies
approximately like $T^{-1}$ at that time. This will prove to
be helpful when we derive an approximate formula for the relic density
$\omegachi$ in the next section. Numerical results are nevertheless
based on equation~(\ref{a_to_T}).

\vskip 0.1cm
We parametrize the contribution of quintessence to the overall energy
density through
\beq
\etaphi \; = \;
{\displaystyle \frac{\rho_{\Phi}^{0}}{\rho_{\gamma}^{0}}} \;\; .
\eeq
The ratio $\etaphi$ of the quintessence--to--photon energy densities
is defined at temperature $T_{0}$. The latter has been set equal to 1 MeV.
Because quintessence should not upset the conventional BBN scenario, we
expect in principle $\etaphi \le 1$. Note however that a non--vanishing
value for $\etaphi$ may still alter the pre--BBN era because quintessence
may be the dominant form of energy at that time if kination holds. Its
energy density varies actually like
\beq
{\displaystyle \frac{\rho_{\Phi}}{\rho_{\gamma}^{0}}} \; = \;
\etaphi \, \left\{
{\displaystyle \frac{h_{\rm eff}(T)}{h_{\rm eff}(T_{0})}} \right\}^{2} \,
\left( {\displaystyle \frac{T}{T_{0}}} \right)^{6} \;\; ,
\eeq
whereas for radiation, the temperature dependence is smoother with
\beq
{\displaystyle \frac{\rho_{\rm rad}}{\rho_{\gamma}^{0}}} \; = \;
g_{\rm eff}(T) \,
\left( {\displaystyle \frac{T}{T_{0}}} \right)^{4} \;\; .
\eeq
Going back in time, $\rho_{\Phi}$ rises more steeply than $\rho_{\rm rad}$.
Even for small values of $\etaphi$, quintessence dominates over the radiation at
early times. Up to a numerical constant, we conclude that the expansion rate $H$
is increased by a factor of $\sqrt{\etaphi} \, \left( T / T_{0} \right)$ with
respect to the conventional radiation dominated cosmology. The next section is
devoted to the consequences of such a change in $H$ on the decoupling of
neutralinos.

\section{The freeze--out of neutralinos.}
\label{sec:decoupling}

At early times -- as long as the temperature $T$ exceeds
their mass $m_{\chi}$ -- neutralinos are in chemical equilibrium.
They steadily annihilate into lighter species while the reverse
process is concomitantly active. The annihilation--production
reaction
\beq
\chi  + \overline{\chi} \; \rightleftharpoons \;
f     \, \overline{f} \; , \;
W^{+} \, W^{-} \; , \;
Z^{0} \, Z^{0} \; , \;
H     \, H     \;   \ldots
\label{salati:REAC2}
\eeq
reaches its thermodynamical equilibrium. This implies a neutralino
density
\beq
n_{\chi}^{0} \; = \; g_{\chi} \; T^{3} \; e^{\displaystyle - y} \;
\left( \frac{y}{2 \, \pi} \right)^{3/2} \;\; ,
\label{salati:DENSITY_NR}
\eeq
that only depends on the mass--to--temperature ratio $y = {m_{\chi}}/{T}$
as well as on the number $g_{\chi}$ of spin states.
As soon as $T$ drops below $m_{\chi}$ as the result of the overall
adiabatic expansion, the neutralino population becomes non--relativistic
and the annihilations take over the thermal productions. Neutralinos
are severely depleted until their density $n_{\chi}^{0}$ is so low that
they fail to annihilate with each other. Reaction (\ref{salati:REAC2}) stops
to be in equilibrium. The resulting freeze--out occurs typically for values
of the mass--to--temperature of $y_{F} \sim 20$. Because the probability
for a neutralino to encounter a partner has become less than unity per
typical expansion time $H^{-1}$, the density $\nA$ remains subsequently
constant per covolume -- per volume that expands with the expanding
universe. The relic abundance $\omegachi$ readily obtains from the value
of $n_{\chi}^{0}$ at freeze--out.

\begin{figure*}[!h]
\centerline{\epsfig{file=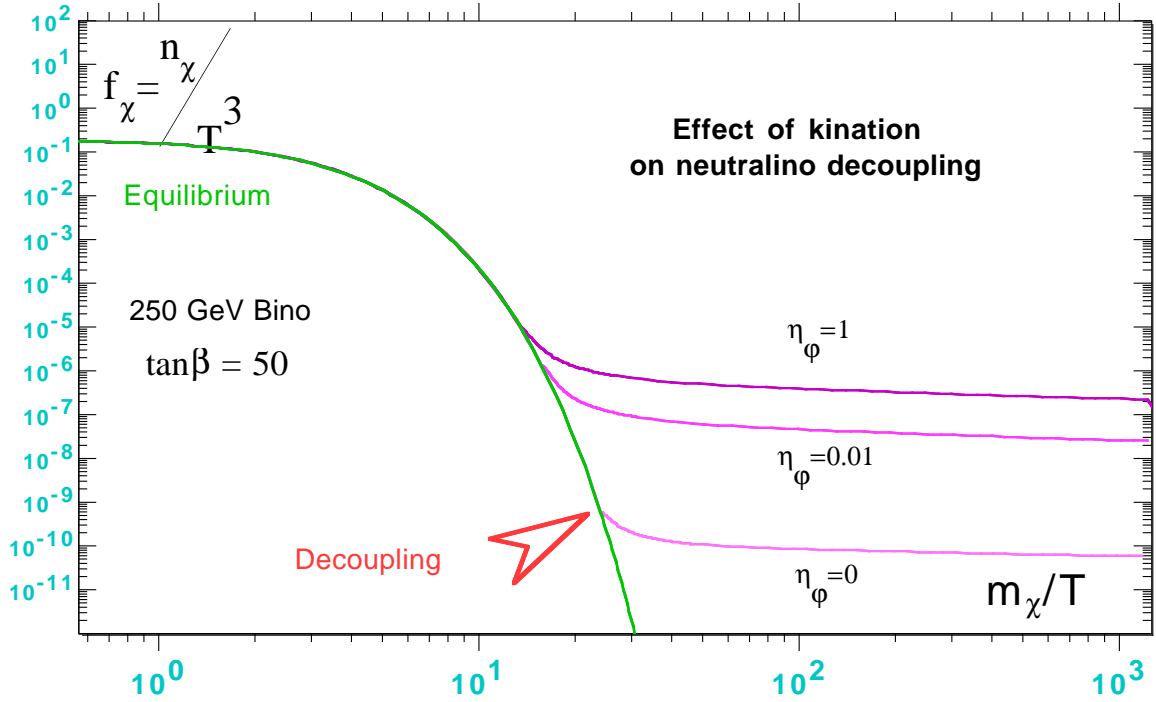,width=0.85\textwidth}}
\vskip 0.5cm
\caption{
Neutralino codensity as a function of the mass--to--temperature
ratio $y = {m_{\chi}}/{T}$ for three different values of the
kination parameter. For $\etaphi = 0$, we recover the standard
radiation dominated cosmology whereas for $\etaphi = 0.01$ and
$\etaphi = 1$, the expansion rate $H$ is significantly increased.
This leads to an earlier decoupling and to a much larger asymptotic
value for the neutralino codensity.
}
\label{fig:FIG_1}
\end{figure*}

\vskip 0.1cm
Assuming that there are as many species $\chi$ than antiparticles
$\overline{\chi}$ -- which is an obvious statement if neutralinos are Majorana
fermions as is the case for instance in the supersymmetric extensions of
the standard model -- the density $\nA$ evolves according to the basic
equation
\beq
\frac{d \nA}{dt} \; = \; - 3 \, H \, \nA \; - \;
< \Sa v > \, \nA^{2} \; + \; < \Sa v > \, \nAe^{\, 2} \;\; .
\label{BOLTZMANN}
\eeq
The first expression in the right--hand side refers to the dilution
resulting from the expansion of the universe. The second term accounts
for the neutralino annihilations. The last expression -- for which detailed
balance has been assumed -- describes the back--creations of $\chi \overline{\chi}$
pairs from lighter species. The neutralino density is given by $\nAe$ as long as
a thermodynamical equilibrium is reached for reaction~(\ref{salati:REAC2}).
In the non--relativistic regime at stake here, it is given by
relation~(\ref{salati:DENSITY_NR}). In terms of the codensity $\fA = \nA / T^{3}$,
the evolution equation simplifies into
\beq
\frac{d \fA}{dt} \; + \; \left\{ < \Sa v > \nA \right\} \, \fA
\; = \; < \Sa v > \, T^{3} \, \fAe^{\, 2} \;\; .
\label{salati:EQUACOD}
\eeq
As discussed in the previous section, the temperature $T$ varies roughly as
the inverse of the scale factor $a$. The codensity $\fA$ corresponds therefore
to the number of particles inside a typical volume $\propto a^{3}$ that 
follows the expansion of the universe. In order to solve
equation~(\ref{salati:EQUACOD}), two typical time scales may be defined.
%
To commence, as long as reaction~(\ref{salati:REAC2}) has reached its equilibrium,
the time derivative $d \fA / dt$ may be neglected and we recover $\fA = \fAe$
as we should. The characteristic time scale with which the neutralino codensity
$\fA$ relaxes towards its kinetic -- and thermodynamical -- equilibrium value
$\fAe$ is set by the annihilation rate
\beq
\tau_{rel}^{-1} \; = \; < \Sa v > \nA \;\; .
\eeq
%
Then, the time scale of the variations of the equilibrium $\fAe$ itself
may be expressed as
\beq
\tau_{eq}^{-1} \; = \; - \frac{d}{dt} \, {\rm Log} \,
\left\{ \fAe^{\, 2} T^{3} \right\} \; \simeq \; 2 \, H \,
\left\{ {\displaystyle \frac{m_{\chi}}{T}} \right\} \;\; ,
\eeq
in the non--relativistic regime.
As is clear in Fig.~\ref{fig:FIG_1}, the neutralino freeze--out proceeds in
two stages.

\begin{figure*}[!h]
\centerline{\epsfig{file=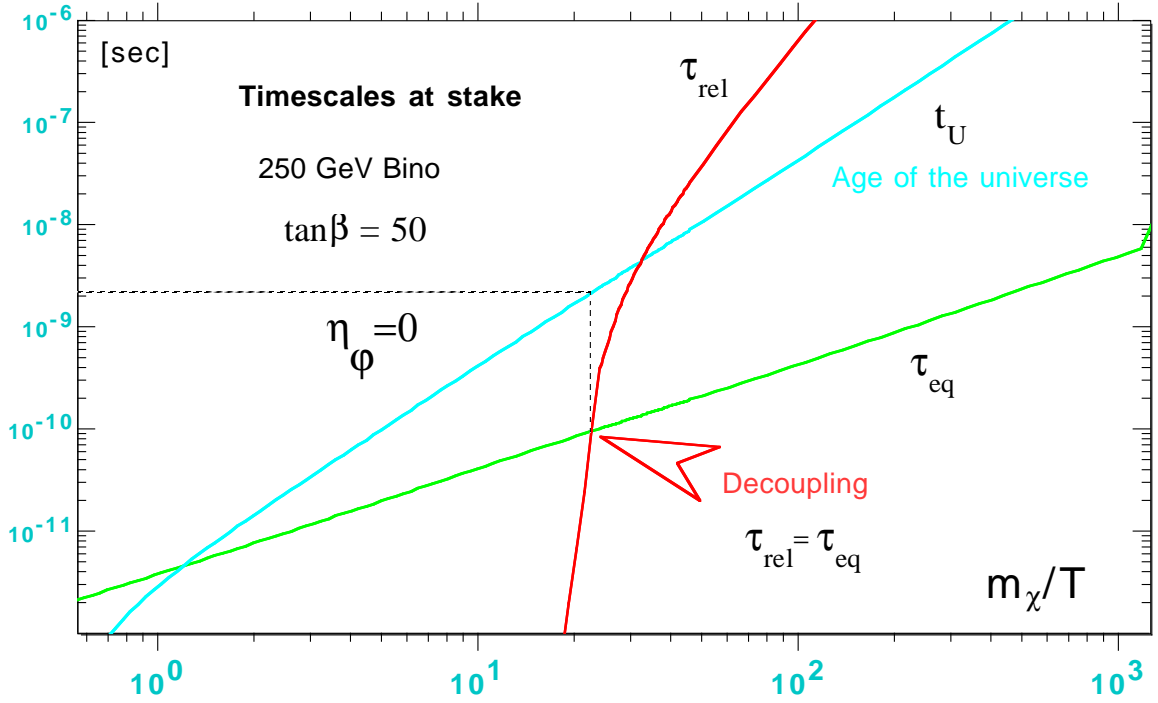,width=0.85\textwidth}}
\vskip 0.5cm
\caption{
The age of the universe $t_{U}$ as well as the typical time
scales $\tau_{rel}$ and $\tau_{eq}$ are featured as a function
of the  mass--to--temperature ratio $y = {m_{\chi}}/{T}$. The
freeze--out occurs at $\yF = 22.7$ when $\tau_{rel}$ overcomes
$\tau_{eq}$. The standard radiation dominated cosmology is assumed
here with $\etaphi = 0$ so that $t_{U}$ evolves like $y^{2}$.
}
\label{fig:FIG_2}
\end{figure*}

\begin{figure*}[!h]
\centerline{\epsfig{file=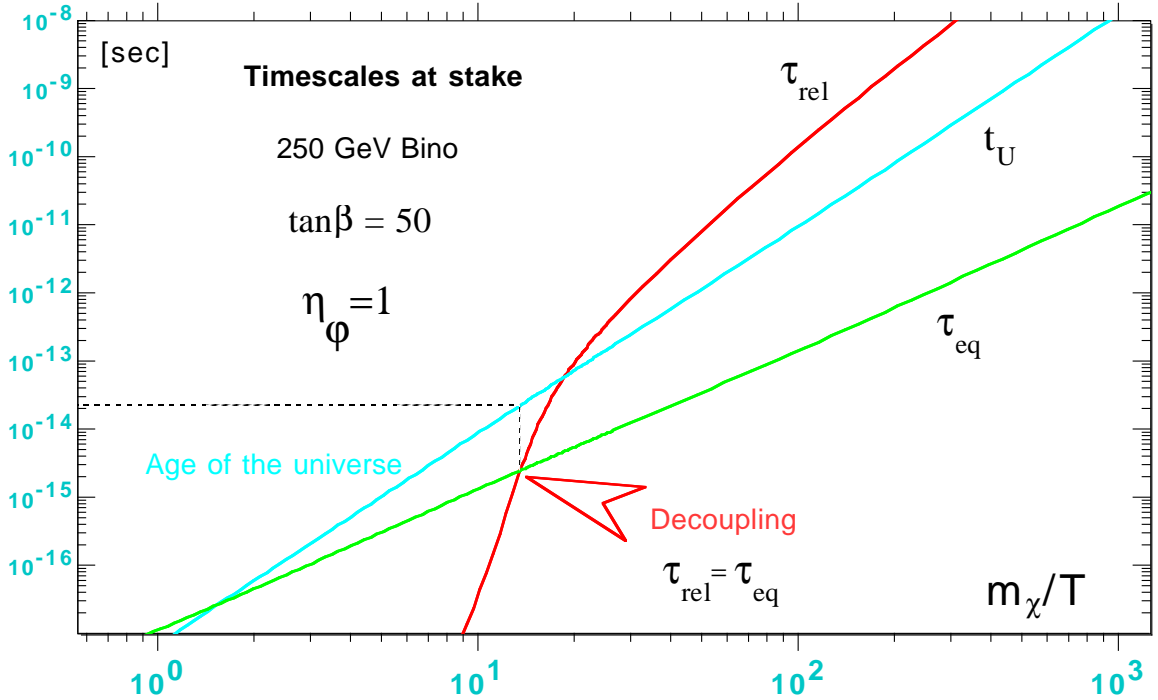,width=0.85\textwidth}}
\vskip 0.5cm
\caption{
Same as in Fig.~\ref{fig:FIG_2} with a kination parameter of
$\etaphi = 1$. The scalar field dominates over the radiation and
the expansion is accelerated with respect to the conventional
situation. This implies smaller values for the age $t_{U}$ that
evolves now like $y^{3}$. The freeze--out point is therefore
reached earlier with $\yF = 13.5$, hence a larger neutralino
relic density.
}
\label{fig:FIG_3}
\end{figure*}

\vskip 0.1cm
\noindent {\bf (i)}
At high temperature, as long as $\tau_{rel} \ll \tau_{eq}$, $\fA$ has
plenty of time to relax towards the equilibrium value $\fAe$ that evolves
at a much slower pace during this stage. As a consequence, the annihilation
reaction~(\ref{salati:REAC2}) reaches thermodynamical equilibrium.
A very good approximation for the neutralino density is provided by $\fAe$ as
featured in Fig.~\ref{fig:FIG_1} where the various neutralino density curves all
merge for $y \lsim 10$ whatever the value of the kination parameter $\etaphi$.
As illustrated by Fig.~\ref{fig:FIG_2} and \ref{fig:FIG_3}, relaxation becomes
progressively less efficient as long as the temperature decreases. The
freeze--out takes place at the very moment when $\tau_{rel}$ crosses $\tau_{eq}$.

\vskip 0.1cm
\noindent {\bf (ii)}
From that moment on -- below the freeze--out temperature $T_{F}$ -- the
relaxation time $\tau_{rel}$ well exceeds $\tau_{eq}$. Whilst $\fAe$ drops
down and vanishes, $\fA$ still decreases a little bit and reaches an asymptotic
value of $\fAasy$ that leads directly to the relic abundance $\Ochi$.

\vskip 0.1cm
To illustrate our discussion, we have considered a generic bino--like
species with mass $m_{\chi} = 250$ GeV and $\tan \beta = 50$. The thermally
averaged annihilation cross section has been approximated by
\beq
< \Sa v > \; \simeq \; \tilde{a} \, + \, \tilde{b} \, x \;\; ,
\label{APPROX_SV}
\eeq
where the parameter $x$ stands for the temperature--to--mass ratio
$T / m_{\chi}$. We have chosen the values
%
$\tilde{a} = 0.1$ pb and $\tilde{b} = 21$ pb for the annihilation cross
section. Before freeze--out, the neutralino density $\nA$ is set equal to
$\nAe$. As soon as decoupling has taken place, we integrate the Boltzmann
equation~(\ref{BOLTZMANN}) in terms of the codensity $\FA = \fA / \kappa$
where the coefficient $\kappa(T) = h_{\rm eff}(T) / h_{\rm eff}(T_{0})$
accounts for the reheating of the radiation as its massive species annihilate
and are converted into lighter populations. As is clear from equation~(\ref{a_to_T}),
the codensity $\FA$ evolves exactly as $a^{-3}$. For convenience, we have
plotted $\fA$ instead of $\FA$ as a function of the $m_{\chi} / T$ ratio
in Fig.~\ref{fig:FIG_1} for three different values of $\etaphi$.
In Figs.~\ref{fig:FIG_2} and \ref{fig:FIG_3}, the age of the universe $t_{U}$
as well as the typical time scales $\tau_{rel}$ and $\tau_{eq}$ are also featured
as a function of the  mass--to--temperature ratio ${m_{\chi}}/{T}$. The
kination parameter $\etaphi$ has been respectively set equal to $0$ and $1$.
When the scalar field dominates over the radiation -- see Fig.~\ref{fig:FIG_3} --
the expansion of the universe is accelerated with much smaller values for $t_{U}$.
The evolution of $\tau_{rel}$ with $y$ is not affected by kination. On the
contrary, $\tau_{eq}$ is much smaller than in the conventional radiation 
dominated cosmology. Freeze--out is therefore reached at a higher temperature
with $\yF = 13.5$ for $\etaphi = 1$ instead of $\yF = 22.7$ when $\etaphi = 0$.
The decoupling temperature increases by a factor of $\sim 2$.
This mild variation nevertheless implies a strong rise in the neutralino
relic density as a result of the strong dependence of $\nAe$ on $y$ -- see
relation~(\ref{salati:DENSITY_NR}). Fig.~\ref{fig:FIG_1} features a growth
of the asymptotic codensity $\fAasy$ by a factor of $\sim$ 350 when $\etaphi$
is varied from 0 to 0.01. When $\etaphi$ is set equal to 1, that increase
reaches a factor of $\sim$ 2900.

\vskip 0.1cm
The previous example illustrates the large increase of the neutralino
relic abundance when kination takes over radiation in the pre--BBN
period. The effect of $\etaphi$ on $\Ochi$ may also be understood
in the framework of the simple approximation which we derive now.
Integrating the cross section $\Sa V$ over the thermal distribution of
neutralinos is beyond the scope of this work. We will simply assume here
that relation~(\ref{APPROX_SV}) applies. This is certainly correct as long
as the annihilation does not proceed through a s--channel resonance for
which the presence of a pole implies an unusual behaviour of the cross
section and we will postpone this problem to a later analysis.
The freeze--out occurs at temperature $\TF$ when $\tau_{rel} = \tau_{eq}$.
This condition translates into
\beq
\left\{ \tilde{a} \, + \, \tilde{b} \, \xF \right\} \,
\TF^{3} \, \fF \; = \;
2 \, \yF \, H \left( \TF \right) \;\; .
\label{F_CONDITION}
\eeq
At temperature $T$, the expansion rate $H$ is increased with respect to
the conventional situation by a factor of $\sqrt{1 \, + \, \alpha \, x^{2}}$
where $x = T / m_{\chi}$. The parameter $\alpha$ describes the relative
contribution of quintessence to the overall energy density as compared to the
radiation
\beq
\alpha \; = \;
{\displaystyle \frac{\etaphi}{g_{\rm eff}(T)}} \,
\left\{ {\displaystyle \frac{m_{\chi}}{T_{0}}} \right\}^{2} \,
\kappa^{2} \;\; .
\eeq
%
The evolution of $\fA$ after decoupling is followed by integrating
relation~(\ref{salati:EQUACOD}) from $x = \xF$ down to $x = 0$
while neglecting the right--hand side term. This leads to a decrease of
the codensity $\fA$ just after freeze--out from $\fF$ down to its
asymptotic value $\fAasy$ with
\beq
\fAasy \; = \;
{\displaystyle \frac{\fF}{\mu}} \;\; .
\label{DEFINITION_MU}
\eeq
The decrease factor $\mu$ depends on $\xF$ and on the parameter $\alpha$
which we evaluate at the decoupling temperature $\TF$. This assumption is
reasonable because most of the evolution of $\fA$ takes place immediatly after
freeze--out as featured in Fig.~\ref{fig:FIG_1}. We readily obtain
\beq
\mu \left( \xF , \alpha \right) \, - \, 1 \; = \;
{\displaystyle \frac{2}{\xF}} \,
\sqrt{1 \, + \, \alpha \, \xF^{2}} \,
\left\{ {\displaystyle \frac
{\tilde{a} \, A \, + \, \tilde{b} \, \xF \, B}
{\tilde{a}      \, + \, \tilde{b} \, \xF     }} \right\} \;\; ,
\eeq
where the factors $A$ and $B$ depend on the combination
$u = \sqrt{\alpha} \, \xF$ through the relations
\beq
A(u) \; = \;
{\displaystyle \frac
{\ln \left\{
{u \, + \, \sqrt{\displaystyle 1 \, + \, u^{2}}} \right\}}{u}} \;\;
\eeq
and
\beq
B(u) \; = \;
{\displaystyle \frac
{\sqrt{\displaystyle 1 \, + \, u^{2}} \, - \, 1}{u^{2}}} \;\; .
\eeq
In the conventional scenario for which $\alpha = 0$, the previous
expressions simplify into $A(0) = 1$ and $B(0) = 1/2$.
%
%
The freeze--out codensity may be derived from the condition~(\ref{F_CONDITION})
\beq
\fF \; = \;
\sqrt{\displaystyle \frac{32 \, \pi^{3}}{45}} \,
\left\{ g_{\rm eff}(\TF) \right\}^{1/2} \,
{\displaystyle \frac{\sqrt{1 \, + \, \alpha \, \xF^{2}}}{\xF^{2}}} \,
\left\{ m_{\chi} \, M_{\rm P} \,
\left( \tilde{a} \, + \, \tilde{b} \, \xF \right) \right\}^{-1} \;\; ,
\eeq
where $M_{\rm P}$ denotes the Planck mass. The asymptotic codensity
$\fAasy$ is defined by relation~(\ref{DEFINITION_MU}) and we infer
for the present epoch a neutralino relic density of
\beq
\rho_{\chi} \; = \;
\frac{4}{11} \, {\displaystyle \frac{1}{\kappa(\TF)}} \; T_{\gamma}^{3} \;
\fAasy \, m_{\chi} \;\; .
\eeq
The factor $(4 / 11) \, (1 / \kappa)$ accounts for the reheating of the
photon background that occurs in the period extending from the neutralino
freeze--out until now. Most of the relevant dark matter candidates have a
mass $m_{\chi}$ in the range between 100 GeV and 10 TeV. Notice also that
the freeze--out temperature--to--mass ratio $\xF \sim 0.1 - 0.2$ is not
very sensitive to the quintessence parameter $\etaphi$. We can safely take
$g_{\rm eff}(\TF) \simeq h_{\rm eff}(\TF) \simeq 45$ for a numerical
estimate of the previous expression in order to get
\beq
\rho_{\chi} \; \simeq \;
0.69 \; {\rm keV \; cm^{-3}} \;\;
{\displaystyle \frac{\sqrt{1 \, + \, \alpha \, \xF^{2}}}{\mu \, \xF^{2}}} \,
\left\{ {\displaystyle \frac
{3 \times 10^{-27} \; {\rm cm^{3} \; s^{-1}}}{< \Sa v >}} \right\} \;\; .
\eeq
That result is to be compared to the closure density
\beq
\rclose \; = \; 1.879 \times 10^{-29} \; {\rm g \; cm^{-3}} \; h^{2}
\; \simeq \; 10.6 \; {\rm keV \, cm^{-3}} \; h^{2} \;\; ,
\eeq
hence a neutralino relic abundance that may be expressed as
\beq
\Ochi \; \simeq \; 6.6 \times 10^{-2} \;\;
{\displaystyle \frac{\sqrt{1 \, + \, \alpha \, \xF^{2}}}{\mu \, \xF^{2}}} \,
\left\{ {\displaystyle \frac
{3 \times 10^{-27} \; {\rm cm^{3} \; s^{-1}}}{< \Sa v >}} \right\} \;\; ,
\label{OMH2_APRX}
\eeq
where the annihilation cross section is taken at decoupling and may
differ from its present value if $\tilde{b}$ well exceeds $\tilde{a}$.
The approximation~(\ref{OMH2_APRX}) tends to overestimate the relic abundance
but is still acceptable. In our generic example illustrated in the previous
figures, we have numerically derived a value of $\Ochi = 0.145$ for $\etaphi = 0$
and of $\Ochi = 440$ for $\etaphi = 1$ whereas expression~(\ref{OMH2_APRX})
respectively gives $\Ochi = 0.16$ and 465.


\section{Discussion and prospects.}
\label{sec:discussion}

\vskip 0.1cm
We have shown that the neutralino relic abundance increases if a period
of kination takes place during the freeze--out of the species. We derive
now an estimate of the corresponding boost factor as a function of $\etaphi$
and $m_{\chi}$.
If $\tilde{a}$ dominates over $\tilde{b} \, \xF$ in the expression
of the annihilation cross section, we may even simplify further
relation~(\ref{OMH2_APRX}) in order to get
\beq
\Ochi \; \sim \; \left\{ {\displaystyle \frac
{2 \times 10^{-27} \; {\rm cm^{3} \; s^{-1}}}{\tilde{a}}} \right\} \;\;
\eeq
in the conventional radiation dominated cosmology. We have taken a
value of $\yF \sim 20$ for the mass--to--decoupling temperature ratio
in that case. On the contrary, if quintessence is the dominant form
of energy with a large value for $\etaphi$, the neutralino fossile
abundance becomes
\beq
\Ochi \; \sim \;
{\displaystyle \frac{\sqrt{\etaphi} \; m_{100}}
{\ln \left( 2 \, u \right)}} \, \left\{ {\displaystyle \frac
{1.3 \times 10^{-23} \; {\rm cm^{3} \; s^{-1}}}{\tilde{a}}} \right\} \;\; ,
\eeq
where $u = \sqrt{\alpha} \, \xF$ has already been defined. With a value
of $\yF \sim 10$ -- see the previous section -- this implies
\beq
u \; \simeq \; 1.5 \times 10^{4} \; \sqrt{\etaphi} \; m_{100} \;\; ,
\eeq
so that the logarithm yields a contribution $\sim$ 10. The parameter
$m_{100}$ denotes the neutralino mass in units of 100 GeV. A crude
estimate of the relic abundance in this regime ensues
\beq
\Ochi \; \sim \; \sqrt{\etaphi} \; m_{100} \, \left\{ {\displaystyle \frac
{1-2 \times 10^{-24} \; {\rm cm^{3} \; s^{-1}}}{\tilde{a}}} \right\} \;\; .
\eeq
We derive a boost factor of $\sim 10^{3} \, \sqrt{\etaphi} \; m_{100}$
with respect to the conventional cosmology.
If now $\tilde{b} \, \xF$ is the leading term as regards the annihilation
cross section, we find that the relic abundance which is normally given by
\beq
\Ochi \; \sim \; \left\{ {\displaystyle \frac
{10^{-25} \; {\rm cm^{3} \; s^{-1}}}{\tilde{b}}} \right\} \;\; ,
\eeq
is increased to
\beq
\Ochi \; \sim \; \sqrt{\etaphi} \; m_{100} \, \left\{ {\displaystyle \frac
{1.2 \times 10^{-22} \; {\rm cm^{3} \; s^{-1}}}{\tilde{b}}} \right\} \;\; .
\eeq
in the presence of quintessence. The boost factor is still of the order of
$\sim 10^{3} \, \sqrt{\etaphi} \; m_{100}$. In our fiducial illustration,
we actually obtained an increase of the neutralino relic abundance by a factor
of $3,000$ with $m_{100} = 2.5$ and $\etaphi = 1$ in good agreement with
the bench mark value which has been derived here.

\vskip 0.1cm
The increase of $\Ochi$ with $\etaphi$ has interesting consequences
and brings up new perspectives as regards neutralino dark matter.
To commence, the various avatars of the minimal or non--minimal
supersymmetric extensions of the standard model start to be
constrained -- should R parity be conserved -- by the accelerator
data on the one hand side and by the requirement that the neutralino
relic abundance should not overclose the universe or even exceed
the observed value of $\omegaM \, h^{2} = 0.135 \pm 0.009$ \cite{ellis}.
If a period of kination takes place in the pre--BBN period,
the various SUSY configurations in the $(\Ochi , m_{\chi})$ plane
that are so far allowed are shifted upwards with the consequence
of becoming forbidden. Exploring in greater detail this question is
a worthwhile project.

\vskip 0.1cm
We already anticipate that configurations with a very small relic density
-- for instance those for which poles dominate in the annihilation mechanism
-- would become cosmologically attractive if $\etaphi$ is large
enough. The difference with the conventional cosmology lies in the
significant enhancement of the annihilation cross section of neutralino
dark matter candidates. At fixed $\Ochi$, notice
that $\Sa v$ increases precisely by the same factor of
$\sim 10^{3} \, \sqrt{\etaphi} \; m_{100}$ which we have derived above.
%
%
This means a general enhancement of the various indirect signatures
for supersymmetric dark matter. If neutralinos dominate the mass
budget of the Milky Way halo, they should still annihilate today
and produce gamma--rays, antiprotons and positrons which may be detected
through the corresponding distortions in the various energy spectra.
As a matter of fact, the recent
HEAT experiment has confirmed \cite{heat} an excess around 8 GeV in
the positron spectrum of cosmic rays. A large boost factor of
$\sim 10^{3}$ -- $10^{4}$ in most of the supersymmetric parameter space
is needed to explain that excess in terms of a homogeneous distribution of
annihilating galactic neutralinos. A certain degree of clumpiness is
actually expected in most of the numerical simulations but even in the
extreme case of \cite{moore_clumps}, it does not exceed a few hundreds.
A period of kination in the early universe could provide an alternate
explanation for that boost factor.

\vskip 0.1cm
Another potential consequence of quintessence is the rehabilitation
of a fourth generation heavy neutrino in the realm of the dark matter
candidates. In the conventional cosmology, a 100 GeV stable neutrino
provides today a contribution of $\sim 10^{-4}$ to the closure density.
Once again, kination at the time of decoupling would enhance that
relic density and make that species cosmologically relevant.

\vskip 0.1cm
Notice finally that scenarios with extra--dimensions have the same effect
as kination. The expansion rate is also increased and may even evolve
as $T^{4}$ -- to be compared to a $T^{3}$ behaviour in our case and to
a $T^{2}$ dependence in the conventional radiation dominated universe.
Implications of such scenarios on neutralino dark matter should be
investigated.
In the case of a low reheating temperature at the end of inflation,
neutralinos are not thermally produced. Depending on the details of the
scenario -- decay of an inflaton field \cite{giudice} or on the contrary
decay of moduli fields \cite{khalil} -- the relic abundance is decreased
or increased.

\vskip 0.1cm
A key ingredient of our study is the contribution $\etaphi$ of the
quintessential scalar field to the overall energy density at the onset of BBN.
A detailed analysis of the light element yields in the presence of kination
\cite{melchiorri} is mandatory at that stage in order to explore a promising
scenario or to derive constraints on $\etaphi$.
%
The existence of dark energy opens up an exciting line of research
and the study of its implications on the astronomical dark matter problem
will certainly bring surprising results.

\section*{Acknowledgements}
\vskip 0.1cm
I would like to thank M.~Joyce for having pointed out to me this
problem in suggesting that a period of kination could potentially
affect the freeze--out of neutralinos and modify their relic density.


\end{document}